%% file: DPF2019_template.tex
\documentclass[11pt]{article}
\usepackage[margin=1in]{geometry}
\usepackage{graphicx}
\usepackage{float}

\input econfmacros.tex 

\def\Title#1{\begin{center} {\Large {\bf #1} } \end{center}}
\def\Author#1{\begin{center} {\normalsize {\sc #1} } \end{center}}
\def\Institution#1{\begin{center} {\normalsize {\it #1} } \end{center}}
\def\Abstract#1{\noindent {\normalsize {\bf Abstract:} {\normalfont #1}}}
\def\Conference{\vspace{4mm}\begin{raggedright} {\normalsize {\it Talk presented at the 2019 Meeting of the Division of Particles and Fields of the American Physical Society (DPF2019), July 29--August 2, 2019, Northeastern University, Boston, C1907293.} } \end{raggedright}\vspace{4mm}}

\begin{document}

%
%

\Title{Electron Neutrino Energy Reconstruction in NOvA Using CNN Particle IDs}

\Author{Shiqi Yu}

\Institution{Argonne National Laboratory\\ Illinois Institue of Technology}

\Abstract{NOvA is a long-baseline neutrino oscillation experiment. It is optimized to measure $\nu_e$ appearance and $\nu_{\mu}$ disappearance at the Far Detector in the $\nu_{\mu}$ beam produced by the NuMI facility at Fermilab. NOvA uses a convolutional neural network (CNN) to identify neutrino events in two functionally identical liquid scintillator detectors. A different network, called prong-CNN, has been used to classify reconstructed particles in each event as either lepton or hadron. Within each event, hits are clustered into prongs to reconstruct final-state particles and these prongs form the input to this prong-CNN classifier. Classified particle energies are then used as input to an electron neutrino energy estimator. Improving the resolution and systematic robustness of NOvA’s energy estimator will improve the sensitivity of the oscillation parameters measurement. This paper describes the methods to identify particles with prong-CNN and the following approach to estimate $\nu_e$ energy for signal events.}

\Conference

%
%

\section{Introduction}

NOvA measures neutrino oscillation parameters by comparing the oscillated neutrino flux at the Far Detector (FD) to the unoscillated flux at the Near Detector (ND). Two oscillation channels are of interest: $\nu_e$ appearance ($\nu_{\mu} \rightarrow \nu_e$) and $\nu_{\mu}$ disappearance ($\nu_{\mu} \rightarrow \nu_{\mu}$). Different parameters can be extracted by using different oscillation probability formula. For example, the probability of $\nu_e$ appearance is

\begin{equation}\label{nueapp}
 P(\nu_{\mu} \rightarrow \nu_e) \approx \left|\sqrt{P_{atmosphere}}e^{-i(\frac{\Delta m^2_{32}L}{4E}+\delta_{CP})}+\sqrt{P_{solar}}\right|^2.
\end{equation}

In Eq.~\ref{nueapp}, $P_{atmosphere}=sin^2\theta_{23}sin^22\theta_{13}sin^2\frac{\Delta m^2_{32}L}{4E}$ is dominated by the $\theta_{13}$ value which is measured by the reactor neutrino experiments, such as the Double Chooz, Daya Bay and RENO reactor neutrino experiments. $P_{solar}$ term contains oscillation parameters, such as $\theta_{12}$ and $\Delta m^2_{21}$, which are measured by solar neutrino experiments such as the Sudbury neutrino observatory and KamLAND experiment.

With an improved neutrino energy reconstruction, NOvA can improve sensitivity to neutrino mass ordering, i.e., $\Delta m_{32}^2$ and provide a stronger constraint on value of $\delta_{cp}$. 

\section{Reconstruction in NOvA detectors.}

NOvA's detectors are made of oriented orthogonal layers, which allow 3D reconstruction by combining the readouts from $xz$ planes and $yz$ planes.

\begin{figure}[!htb]
\begin{center}
\includegraphics[width=11cm]{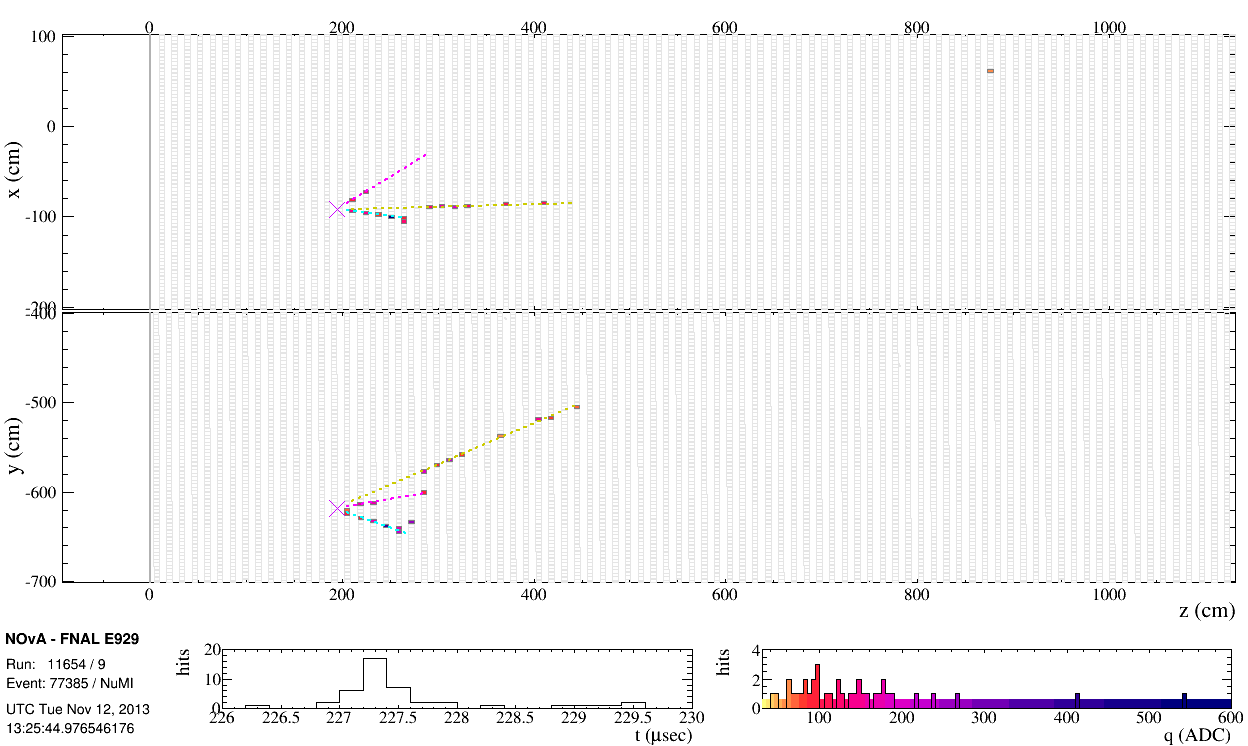}
\caption{Event display with a neutrino candidate in the FD.}\label{event}
\end{center}
\end{figure}
\noindent Figure~\ref{event} shows an example of the $xz$ view and $yz$ view of an event in the FD. With the information from the readouts of two views of each event, an object, which is called slice, is reconstructed based on the time and space correlations of a group of hits. The algorithm is called slicer. Slicer aims at grouping together all the hits belonging to the same neutrino interaction or cosmic event. Ideally, all the hits falling into the same slice come from the same origin, which is the 3D vertex of the event. By minimizing the distance of all the hits to the vertex in angular space, lines, so called Elastic Arms, are drawn and the crossing point is defined as the vertex of the event. The magenta cross in Fig.~\ref{event} is the reconstructed vertex of the event. With the vertex, the following FuzzyK algorithm is applied to reconstruct final-state particles into objects called prongs. A prong contains all the information about a final-state particle's trajectory. There is always at least one prong in each reconstructed event.
\begin{figure}[!htb]
\begin{center}
\includegraphics[width=11cm]{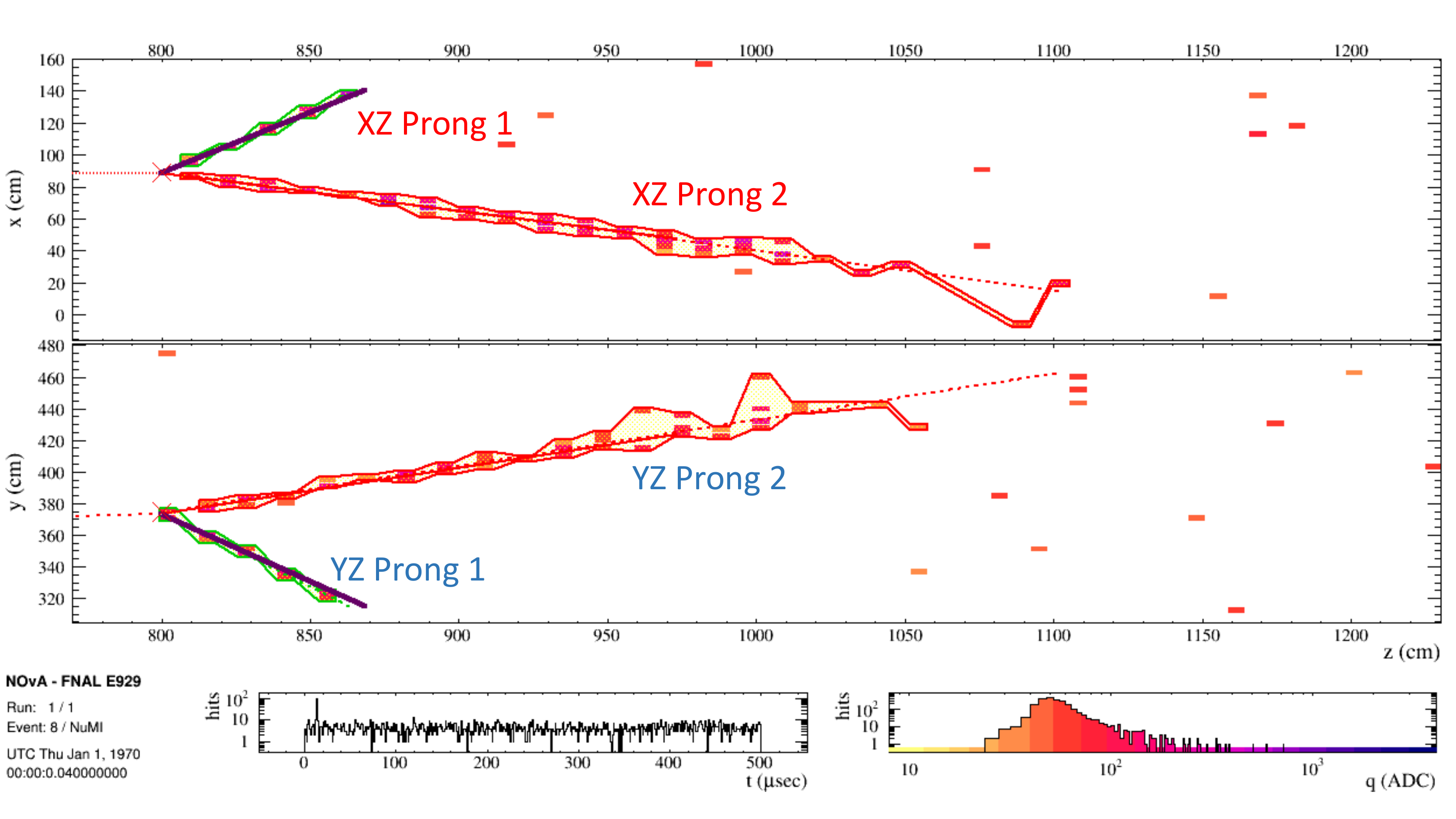}
\caption{Event display of an event in the FD.}\label{eventprongs}
\end{center}
\end{figure}
 
In the $\nu_e$ appearance analysis, the primary out-going leptons detected by the NOvA detectors are electron showers, which have "fuzzy" tracks. FuzzyK performs a clustering algorithm in each view of detector, $xz$ view and $yz$ view, by minimizing the distance among hits, to group hits into clusters, i.e., 2D prongs. An example of clustering is shown in Fig.~\ref{eventprongs}. 

\begin{figure}[!htb]
\begin{center}
\includegraphics[width=8cm]{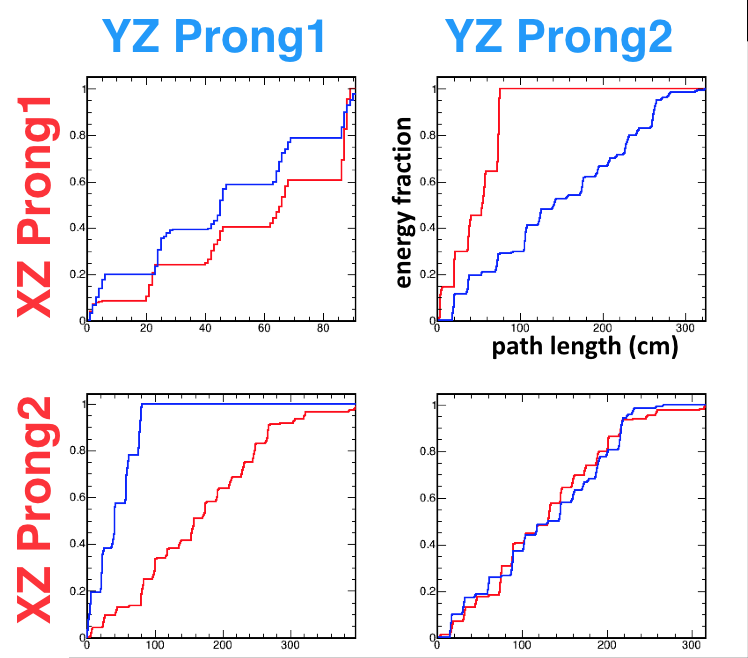}
\caption{View matching of an event in the FD.}\label{matchmatrix}
\end{center}
\end{figure}

Then cumulative energy fraction is calculated along prong length ($\frac{dE}{dx}$), as shown in Fig.~\ref{matchmatrix}. In each event, by comparing the progression of the $\frac{dE}{dx}$ curves, called energy profiles of 2D prongs, FuzzyK attempts to match 2D prongs from each view and reconstruct a pair of ($xz$ view, $yz$ view) 2D prongs into one 3D prong object, i.e., 3D prong~\cite{fuzzyk}.  

\begin{figure}[!htb]
\begin{center}
\includegraphics[width=13cm]{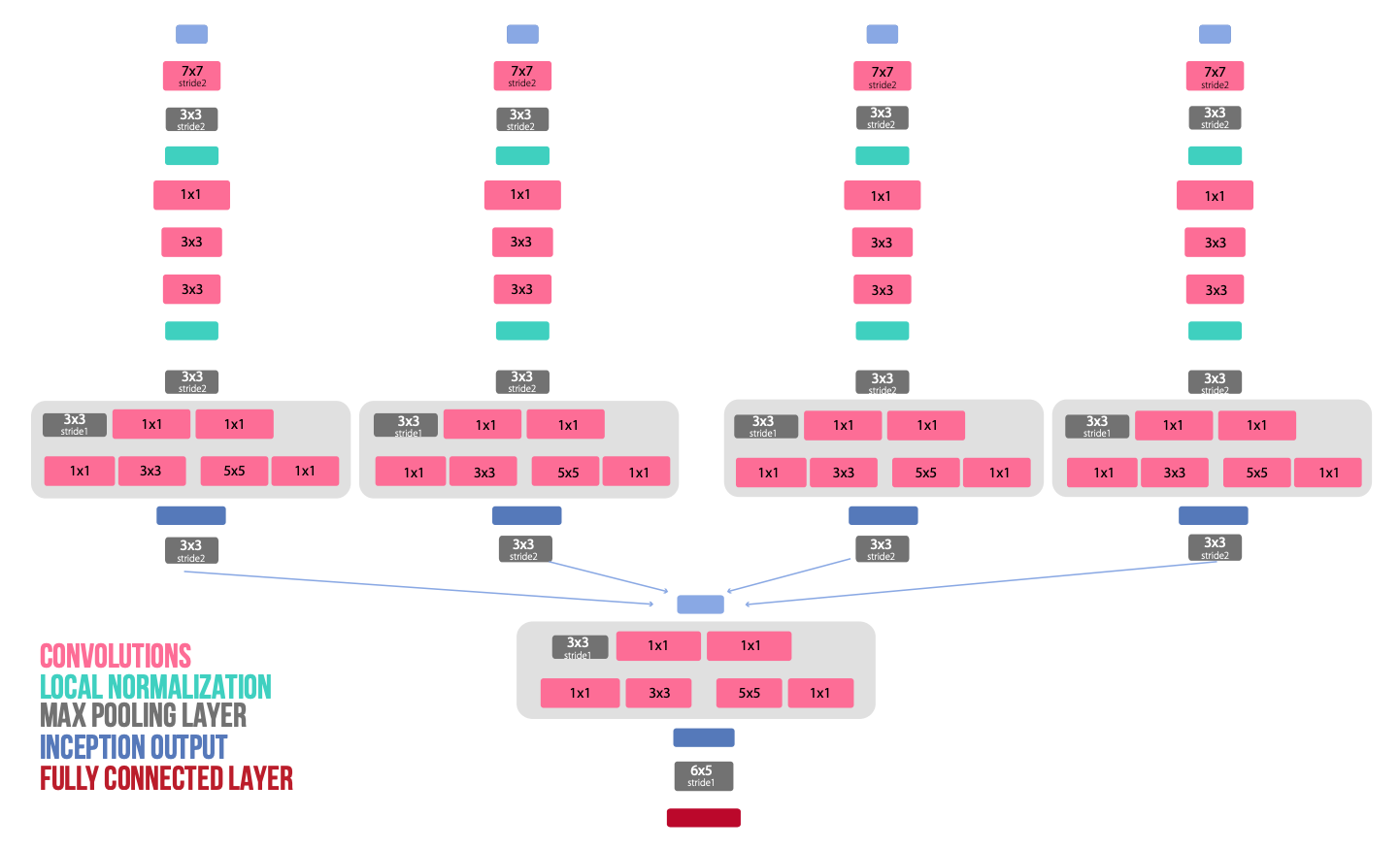}
\caption{Prong-CNN Siamese tower.}\label{prongcvn}
\end{center}
\end{figure}

To get more information on the final-state particles, i.e., 3D prongs, a CNN is employed to classify the 3D prongs' particle types. The input pixel maps of the CNN are the $xz$ and $yz$ views of the 3D prong, and the outputs are the predicted particle ID scores, i.e., the scores for different particle assumptions normalized to unity. The prong-CNN Siamese tower is shown in Fig.~\ref{prongcvn}~\cite{cvnpaper}. \\


\section{Energy Reconstruction}

The approach to the energy reconstruction of $\nu_e$ appearance signal events is based on the assumption that the response of the detectors is inherently different for electromagnetic (EM) and hadronic depositions, both of which are present in the $\nu_e$ appearance signal events. 

By using the output of the prong-CNN classifier, all the 3D FuzzyK prongs are classified as either electromagnetic or hadronic depositions. The method separating them is described as the following. First, two IDs are defined as:

\begin{equation}
CNN_{EM ID} = CNN_{electron\ ID} + CNN_{\gamma \ ID}
\end{equation}
\begin{equation}
CNN_{Hadronic ID} = CNN_{proton \ ID} + CNN_{\pi \ ID} + the\ rest\ IDs.\\
\end{equation}

\vspace{10pt}
\noindent Then, the electromagnetic and hadronic energies are defined based on the prong level information:
\begin{itemize}
    \item EM shower energy ($E_{EM}$): By looping over all the prongs, a prong's calibrated energy is added to $E_{EM}$ if the prong's $CNN_{EM ID}$ is greater than its $CNN_{Hadronic ID}$. 
    \item Hadronic energy ($E_{HAD}$): $E_{HAD}$ = sum of hits energy in the slice - $E_{EM}$, if $E_{HAD}$ $\ge$ 0.
\end{itemize}

\begin{figure}[!htb] 
  \begin{center}
   \includegraphics[angle=0,width=0.6\textwidth]{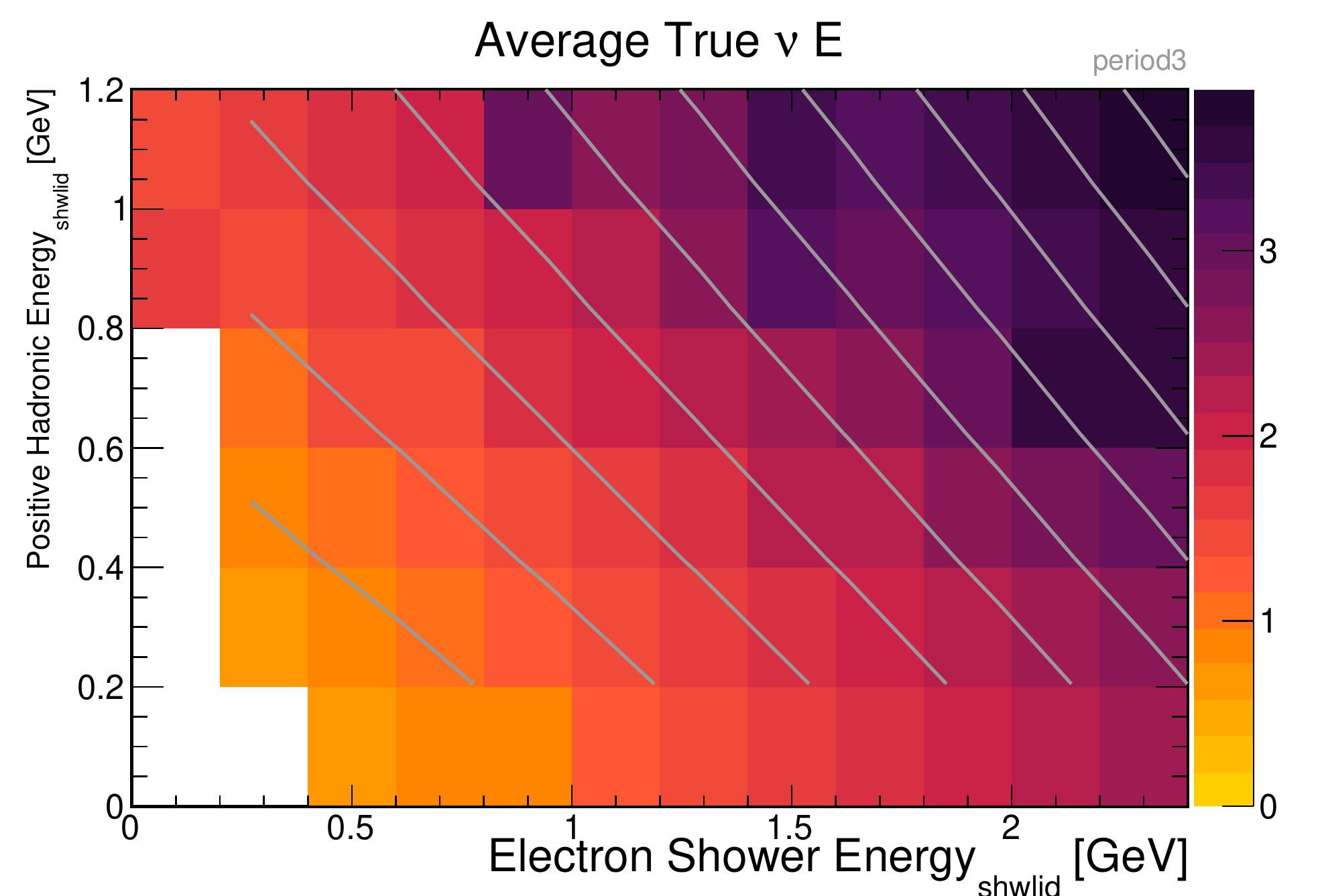}
    \caption{True neutrino energy distribution as a function of reconstructed electromagnetic energy and hadronic energy. Gray line is the contour of weighted average true neutrino energy. }
   \label{fig:FD_makefit}
  \end{center}
\end{figure}

\noindent Once these two components, $E_{EM}$ and $E_{HAD}$, are calculated, they are used as input observables of the $\nu_e$ energy estimator ($E_{reco}$). 

Since NOvA's two detectors are off-axis, the neutrino beam flux is narrowly peaked around 2 GeV. 
In order to keep the energy estimator unbiased with respect to the true energy spectrum, a weight is applied to remove energy dependence from true energy spectral shape. The re-weighed MC is used to fill the 2D spectrum in Fig~\ref{fig:FD_makefit} where x-axis is $E_{EM}$, y-axis is $E_{HAD}$ and each bin is filled by the weighted average true energy value. The gray line shows the contour of weighted averaged true energy. $E_{reco}$ is a function of ($E_{EM}$, $E_{HAD}$). A fit is performed on this 2D spectrum by fitting to the weighted average true energy. Equation~\ref{FHC} is the function that links together $E_{EM}$ and $E_{HAD}$ components. 
\begin{equation}\label{FHC}
E_{reco} = 0.9595 E_{EM} + 1.0198 E_{HAD} +0.0122 E_{EM}^2 +0.2163 E_{HAD}^2 
\end{equation}
This reconstruction $\nu_e$ energy is used in NOvA oscillation analysis\cite{novapaper}.
\section*{Acknowledgements}
I am grateful to Department of Energy for supporting the work.

\end{document}

%% file: econfmacros.tex
\def\beq{\begin{equation}}
\def\eeq#1{\label{#1}\end{equation}}
\def\eeqn{\end{equation}}

\def\beqa{\begin{eqnarray}}
\def\eeqa#1{\label{#1}\end{eqnarray}}
\def\eeqan{\end{eqnarray}}

\let\bar=\overbar

\def\Dslash{\not{\hbox{\kern-4pt $D$}}}
\def\dslash{\not{\hbox{\kern-2pt $\del$}}}

\def\msb{{\bar{\ssstyle M \kern -1pt S}}}

